\documentclass[12pt]{iopart}
\usepackage{amssymb}
\usepackage{graphicx}

\begin{document}

\title{SLE description of the nodal lines of random wave functions}
\author{E. Bogomolny, R. Dubertrand, and C. Schmit}
\address{CNRS, Universit\'e  Paris-Sud, UMR 8626,\\
 Laboratoire de Physique Th\'eorique et Mod\`eles
Statistiques, 91405 Orsay, France}
\ead{remy.dubertrand@lptms.u-psud.fr}

\begin{abstract}
The nodal lines of random wave functions are investigated. We demonstrate
numerically that they are well approximated by the so-called SLE$_6$ curves
which describe the continuum limit of the percolation cluster boundaries. This
result gives an additional support to the recent conjecture that the nodal
domains of random (and chaotic) wave functions in the semi classical limit are
adequately described by the critical percolation theory. It is also shown that
using the dipolar variant of SLE reduces significantly finite size
effects.    
\end{abstract}
\maketitle
\section{Introduction}

In 1977 Berry conjectured \cite{berryrw} that the wave functions of chaotic
quantum systems can statistically be described by Gaussian random functions
with a spectrum computed from the ergodic average. 

For example, the wave functions for two-dimensional billiards without magnetic
field $\Psi(\vec{x}\,)$ have to obey the equation  
\begin{equation}
(\Delta+k^2)\Psi(\vec{x}\,)=0
\label{schro}
\end{equation}
and, say, the Dirichlet boundary conditions. 

From general considerations it follows that such functions can be
represented as a formal sum over elementary solutions of (\ref{schro}) e.g.
\begin{equation}
  \label{form_gen_fc}
  \Psi(r,\theta)=\sum_{m=-\infty}^{\infty} C_m J_{|m|}(k r) {\rm e}^{{\rm i} m \theta}
\end{equation}
where $J_m(x)$ are usual Bessel functions  and $C_{-m}=C_{m}^*$.

The Berry conjecture signifies that for chaotic billiards like the stadium
billiard, real and imaginary parts of coefficients $C_m$  with non-negative
$m$ are independent identically distributed Gaussian random variables with
zero mean and variance computed from the normalization.  

The beauty of this profound conjecture lies first of all in its simplicity and
generality. On the other hand, it represents chaotic wave functions as
completely structureless and 'uninteresting' objects which to some
extent was the reason of relatively few investigations of the chaotic
wave functions.

In 2002 Smilansky {\it et al.} \cite{uzy} stimulated a renewal  of interest in
this problem. Instead of considering the coefficients of an expansion as in
(\ref{form_gen_fc}) these authors looked for the nodal domains of wave functions,
i.e.\ the regions where a function has a definite sign. They have also observed that for
chaotic wave functions the nodal domains have a rich unexpected structure worth
investigating in details.

In \cite{eug_ch_dom_nodal} on physical grounds it was conjectured that the nodal
domains of random wave functions (and, consequently, of chaotic wave
functions) can be adequately 
described by a critical percolation model (see e.g. \cite{stauffer}). 
Due to the universality of the critical percolation it is unessential what special
percolation model is considered. All of them lead to the same  critical
exponents as well as to other universal quantities. In \cite{eug_ch_dom_nodal}
it was checked numerically that many predictions of the critical percolation
model, such as the number of connected domains, their area distribution, cluster 
fractal dimensions etc. agree very well with corresponding quantities computed
from the nodal domains of random (and chaotic) wave functions. In \cite{charles} it was
demonstrated that the level domains (the regions where a function is bigger than a
certain non-zero value) are well described by a non-critical
percolation model. These and other investigations strongly suggest that the
critical percolation model is applicable for the description of the nodal
domains of random and chaotic wave functions. Though it sounds physically
quite natural and can be confirmed by a careful application of Harris'
criterion \cite{charles}, a rigorous mathematical proof was not yet found and further numerical
verifications are desirable.

The purpose of this paper is to check another prediction of the percolation
model namely that the boundaries of the percolation clusters are described by
what is called SLE$_6$ curves (see below). This statement  was proved in
2001 for the critical percolation on the triangular lattice by Smirnov
\cite{smirn} and  it is widely accepted that it remains true for all
critical percolation models. For the nodal 
domains of random functions it leads to the conclusion  that the nodal lines
(curves where a function is zero) have to be also described by SLE$_6$ curves.   

The plan of the paper is the following. In section~\ref{SLE} for
completeness we give an informal introduction to the SLE curves.   
This name is attributed to two-dimensional self-avoiding curves generated by
an one-dimensional Brownian motion with zero mean and the variance linear in
time with a coefficient of proportionality equal to a real positive number
$\kappa$. In 2000 Schramm proved that if the different parts of a random
self-avoiding curve are independent and the curve itself is in a certain
sense conformally  invariant then it belongs to SLE curves with a certain
value of $\kappa$. Smirnov's result \cite{smirn} means that  
the percolation boundaries are generated by the Brownian motion with 
$\kappa=6$. In section~\ref{chordal} we present the results of direct
numerical  calculations of corresponding Brownian-like generating curves for
the nodal lines  of random wave functions and found that they are well described
by SLE$_6$. The inevitable drawback of numerical calculations is that one
always deals with  curves of finite size. But the theorem about  the relation between
the percolation boundaries and SLE$_6$ is valid only for infinite curves. To
decrease such a finite size effect we use in section~\ref{dipolar}  a 
different version of  SLE called dipolar SLE \cite{dipolar}. In this
approach one first conformally transforms a given region into, say, an
infinite  band and then uses a suitably modified SLE equation. By this
method we numerically demonstrate that the nodal lines of
random wave  functions  agree well with the SLE$_6$ description thus once
more confirming the conjectural  relation between the critical percolation
and the nodal domains of random wave functions.  In the appendix the numerical
algorithms used in calculations are shortly discussed. 

To a certain degree our work was stimulated by \cite{turb} where it was
numerically checked that the nodal lines of the vorticity field in
two-dimensional turbulence are close to SLE$_6$ curves. In the recent
publication \cite{kmw}  Keating, Marklof and Williams have demonstrated that
the nodal lines for a perturbed cat map are well described by SLE$_6$. They also
performed \cite{private}  numerical calculations for the nodal lines of random
wave functions which lead to the same conclusions. Our results are in complete
agreement with their findings.

\section{Schramm-Loewner Evolution}\label{SLE}

\begin{figure}[!ht]
  \centering
  \includegraphics[width=.7\linewidth]{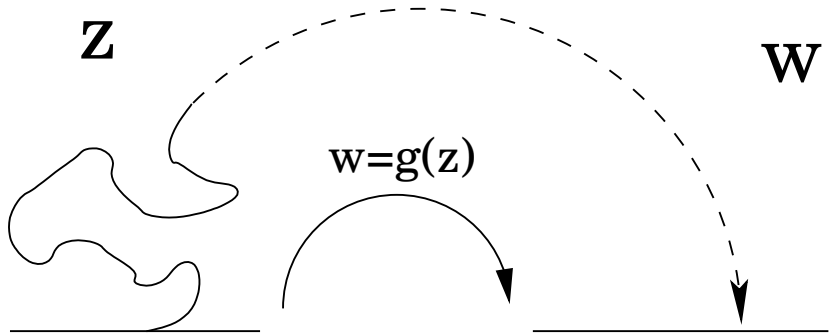}
  \caption{Loewner's evolution. The upper half plane of the variable $z$ with a
    cutoff along a simple curve is transformed by a conformal map $w=g(z)$
    to the whole upper half plane of the variable $w$.} 
   \label{fig1}
\end{figure}

Two-dimensional self-avoiding curves appear naturally in many physically
important problems. But analytically imposing the condition of self-avoiding is not simple.
 In 1923 Loewner proposed \cite{loewner} to describe such curves by
 conformal transformations which map a domain with a simple curve growing
 from the boundary  to another  simply connected domain without the
 curve. In the simplest setting one considers the  upper half plane
 ${\mathbb{H}}$  with a simple curve $\mathcal{C}$ and looks for a conformal
 map which transforms  the upper half plane minus the curve,  
$  {\mathbb{H}} \backslash \mathcal{C}$, to ${\mathbb{H}}$ itself (see figure~\ref{fig1}).

The main point  of this approach is the well known  Riemann theorem
according to which any simply connected region can be conformally mapped to
another simply connected region and, inversely, if one applies a conformal
transformation to a simply connected region the result will be also a simply
connected region.   

Such conformal transformations are not unique and depend on 3 free
parameters. For example, the upper half  plane ${\mathbb{H}}$ is transformed
to itself by 3 parameter group of fractional transformations 
\begin{equation}
g(z)=\frac{a z+b}{c z+d}
\end{equation}
with real $a,b,c,d$.

\begin{figure}[ht]
\begin{center}
\includegraphics[width=.7\linewidth]{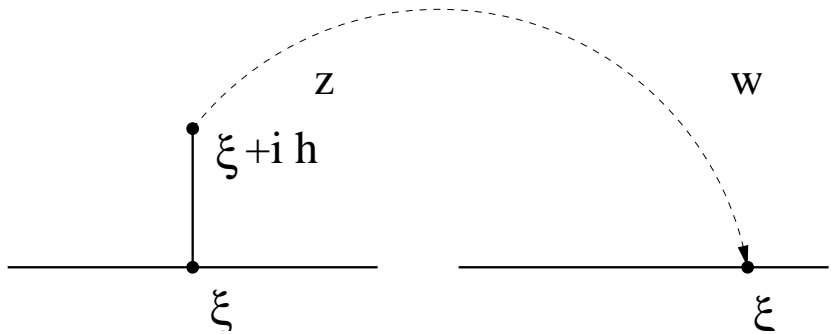}
\end{center}
\caption{Transformation of a vertical slit to the upper half plane by map (\ref{slit}).}
\label{fig2}
\end{figure}

To fix uniquely  the conformal map $g(z)$ it is convenient to impose the so-called 
hydrodynamic normalization by fixing the behaviour of the map at infinity:
\begin{equation}
  g(z) = z+ \frac{2t}{z} +\ldots \textrm{ when } |z| \to \infty\;.
\label{normhy}
\end{equation} 
Parameter  $t$ is a characteristic of the initial region called capacity. To
stress its importance  we shall denote the map $g(z)$ with asymptotic
(\ref{normhy}) by $g_t(z)$ showing explicitly the dependence of $t$.  This
quantity can be also called  'conformal time' as it  fulfils    the
additivity property which  follows directly  from (\ref{normhy}) 
\begin{equation}
 g_t( g_s)=g_{t+s}\;.
 \label{addition}
\end{equation}
 The simplest  example of a curve is a vertical slit in the upper
 half plane. If $(\xi, h)$ are coordinates of the top of the slit (see
 figure~\ref{fig2})  
 the transformation
\begin{equation}
w(z)=\xi +\sqrt{(z-\xi)^2+4t}
\label{slit}
\end{equation}
with $t=h^2/4$ maps the upper half plane of $z$ with the slit to the  upper
half plane of $w$ without the slit.  

Interpreting $t$ as the time permits to derive the equation for $g_t(z)$.
From (\ref{addition}) it follows that  
\begin{equation}
  g_{t+\delta t}(z)=g_{\delta t}\left (g_t(z)\right )\;.
 \end{equation}
For small $\delta t$ any line can approximately be considered as a vertical
slit growing from  the image of the tip of the curve under $g_t$ which we
denote by $\xi_t$  (see appendix).  Now (\ref{slit}) states that  
\begin{equation}
g_{\delta t}(z)\stackrel{\delta t \to 0}{\longrightarrow} z+\frac{2\delta t}{z-\xi_t}\;.
\end{equation}
Combining two last equations  one concludes that  $g_t(z)$ as function of
$t$ obeys the following  differential equation  called Loewner's equation 
\begin{equation}
  \label{loewner}
  \frac{{\rm d} g_t(z)}{{\rm d} t} = \frac{2}{g_t(z)-\xi_t} 
\end{equation}
with initial condition 
\begin{equation}
g_0(z)=z\;.
\end{equation}
The function $\xi_t$ is called the 'driving' or 'forcing' function.  The
curve itself is called the 'trace' and  can be computed from the relation 
\begin{equation}
z( t ) = g_t^{-1}(\xi_t)\;.
\label{trace}
\end{equation}
The advantage of Loewner's equation (\ref{loewner}) and  (\ref{trace})  is
that for any smooth (differentiable)   function $\xi_t$ it will produce a
two-dimensional self-avoiding curve and, inversely, any self-avoiding curve
will give a smooth function $\xi_t$.  

The next important step was done by Schramm. In 2000 he proved
\cite{Schramm} that   random self-avoiding curves with the following
properties: 
\begin{itemize}
\item conformal invariance, 
\item  statistical independence of different parts,
\item  reflection symmetry
\end{itemize}
are generated by the usual one-dimensional  Brownian motion with zero mean
and the variance linear in $t$ 
\begin{equation}
\xi_t=\sqrt{\kappa}B_t\ ,\;\;\;\;<B_t>=0\ ,\;\;\;<B_tB_s>=|t-s|\;.
\label{standard}
\end{equation}
Such curves only depend on one real positive parameter $\kappa$ and are
called  stochastic (or Schramm) -- Loewner evolution curves or shortly
SLE$_\kappa$.   

As Brownian curves are not differentiable, SLE$_\kappa$ curves may have
self-touching points. Depending on the value of $\kappa$ they are  divided  into
three phases  (see e.g. \cite{cardy}).  For $0< \kappa \le 4$ the traces are
simple curves, for $4< \kappa< 8$ they can have double points or hit the
real axis, and for $\kappa\ge 8$ they fill the entire domain. 

SLE is a very powerful tool to study rigorously the scaling (continuum)  limit
of different discrete models (see e.g. \cite{cardy} and \cite{werner}).  
The most important for us is the Smirnov's result \cite{smirn}: the
boundaries of clusters in the critical percolation (on a triangular lattice)
converge in the continuum limit to the traces of SLE$_6$. As the nodal domains of random
wave functions are conjectured to be described by critical percolation
\cite{eug_ch_dom_nodal} it means that the nodal lines of random wave functions
in this limit have to be also described by SLE$_6$ curves. The verification of
this statement is the main purpose of this work. 

\section{The nodal lines of random wave functions}\label{chordal}

\begin{figure}[!ht]
\begin{minipage}[b]{.8\linewidth} 
\includegraphics[angle=-90,width=1.\linewidth]{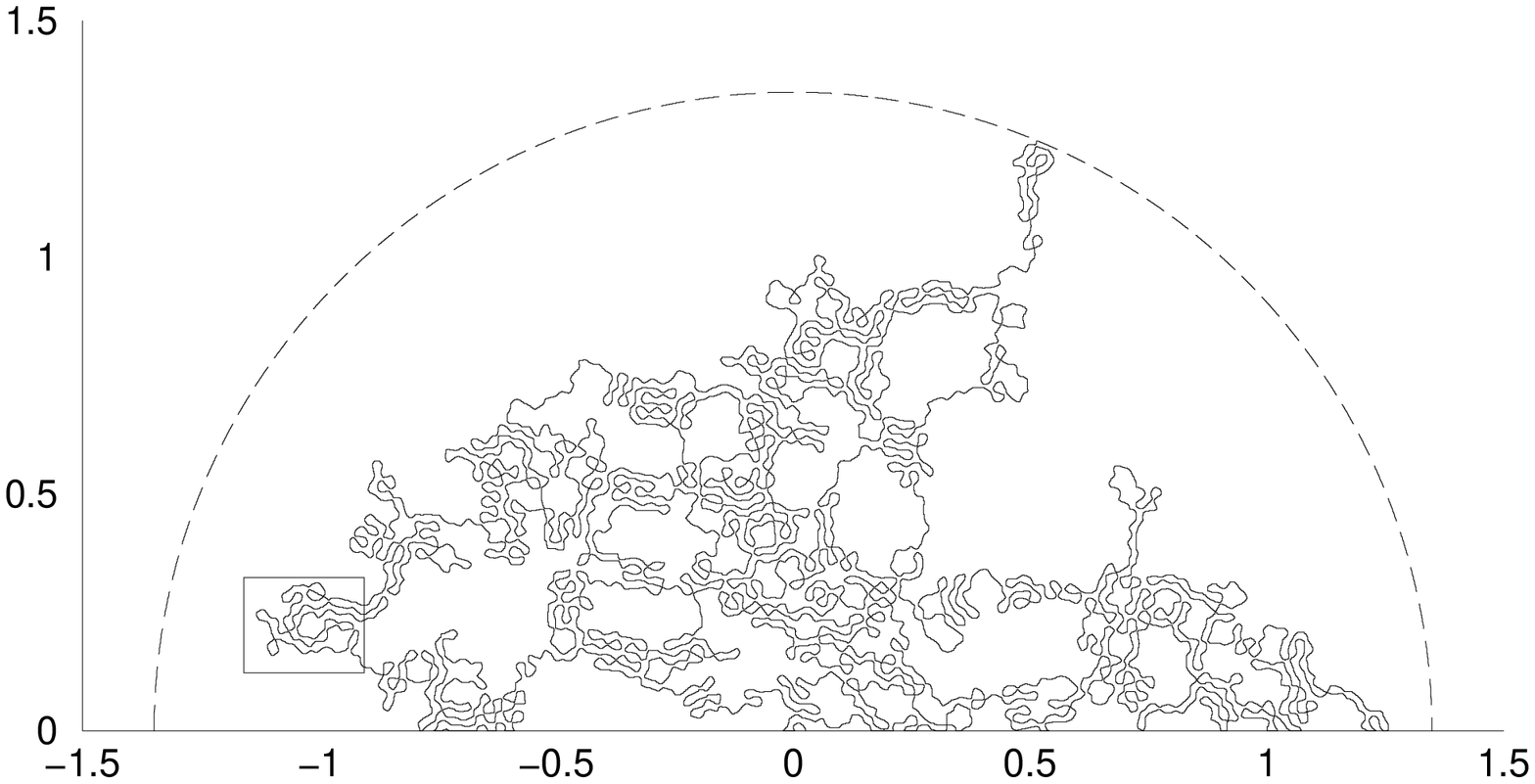}
 \end{minipage}
 \hspace{-.2\linewidth}
 \begin{minipage}[b]{.38\linewidth}
 \centering
 \includegraphics[angle=-90,width=.64\linewidth]{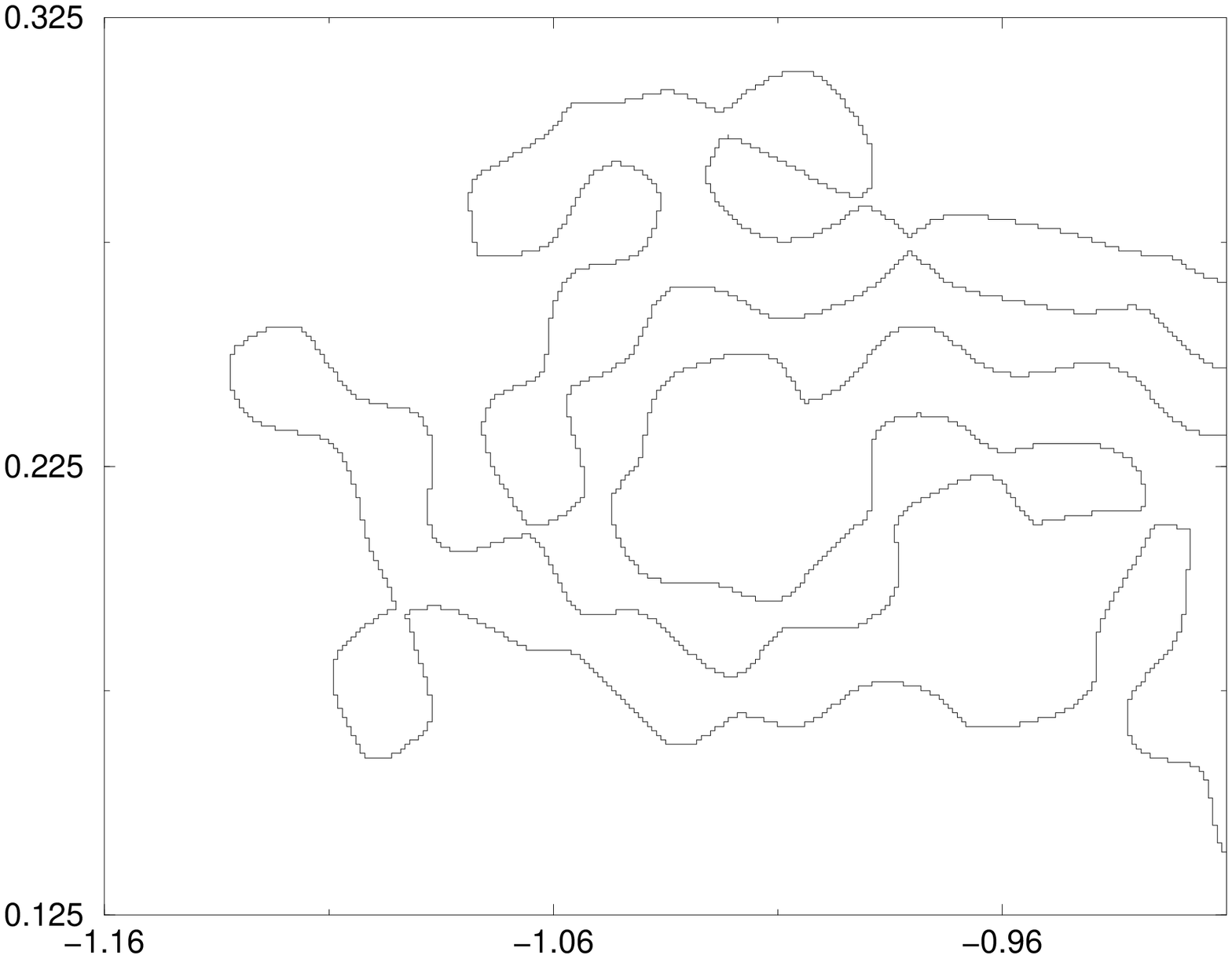}
\end{minipage}
\caption{A nodal line of a random wave function. The dashed line indicates the
  absorbing half-circle. {\it Insert:}  Magnification of the indicated rectangular region. }
\label{exCh}
\end{figure}

The nodal lines are lines where a real function of two variables is zero,
$\Psi(x,y)=0$. They can, in principle, be computed numerically for any
function $\Psi(x,y)$. For random wave functions (\ref{form_gen_fc}) the nodal lines are
quite complicated. An  example of a nodal line is presented in
figure~\ref{exCh}.  To demonstrate the complexity and self-avoiding
character of such a curve, its small portion enclosed by a rectangle is
enlarged in the inset of this figure. 

As all  pictures of nodal lines  are necessarily finite, one has to decide
what to do with a nodal line which hits the boundary. There are two main
types of  boundaries:  reflecting and absorbing. If a growing line touches a
reflecting boundary it follows the boundary in the increasing direction till
the next appearance of  the sign change  and then continues along a new nodal
line. This can be achieved automatically by imposing the sign function outside
the boundary.  If a boundary is absorbing, one simply stops any line which
hits it.  In figure~\ref{exCh} the horizontal line is a reflecting boundary
but the half-circle indicated by the dashed line is an absorbing one.   

In general the properties of  SLE$_\kappa$ curves inside a region do depend on
chosen boundary conditions. But for SLE$_6$ lines it is proved (see
e.g. \cite{cardy}) that they feel the boundaries only when they hit them so the
boundary conditions are unessential for the investigation of the local
properties of the percolation cluster interfaces and, as conjectured, of the nodal
lines of random wave functions.  
\begin{figure}[!h]
      \centering
      \includegraphics[angle=-90,width=.56\linewidth]{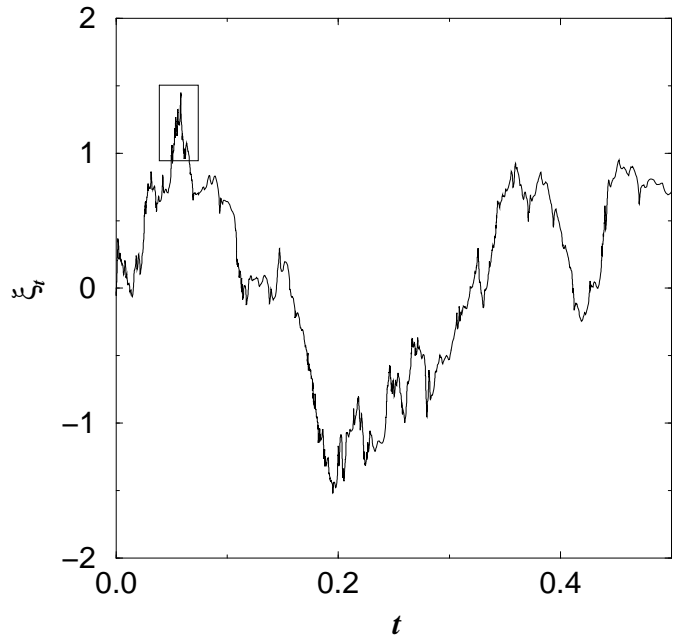}
     \caption{Forcing function corresponding to the nodal line of figure~\ref{exCh}.}
     \label{forcing}
\end{figure}

Having generated a line as in figure~\ref{exCh}, the next step is to find the
conformal map which transforms it to the upper half plane.  There are many good
algorithms which permit to do it numerically (see \cite{algorithm} and
references therein). We used the so-called geodesic algorithm
\cite{algorithm} in which a small segment of a line to be mapped is
approximated by an arc of a geodesic circle perpendicular to the real axis
and passing through the tip of the segment (see appendix). This algorithm is quite stable
and is easy to implement. 

\begin{figure}[!h]     
    \begin{minipage}[b]{.49\linewidth}
      \centering
      \includegraphics[angle=-90,width=.96\linewidth]{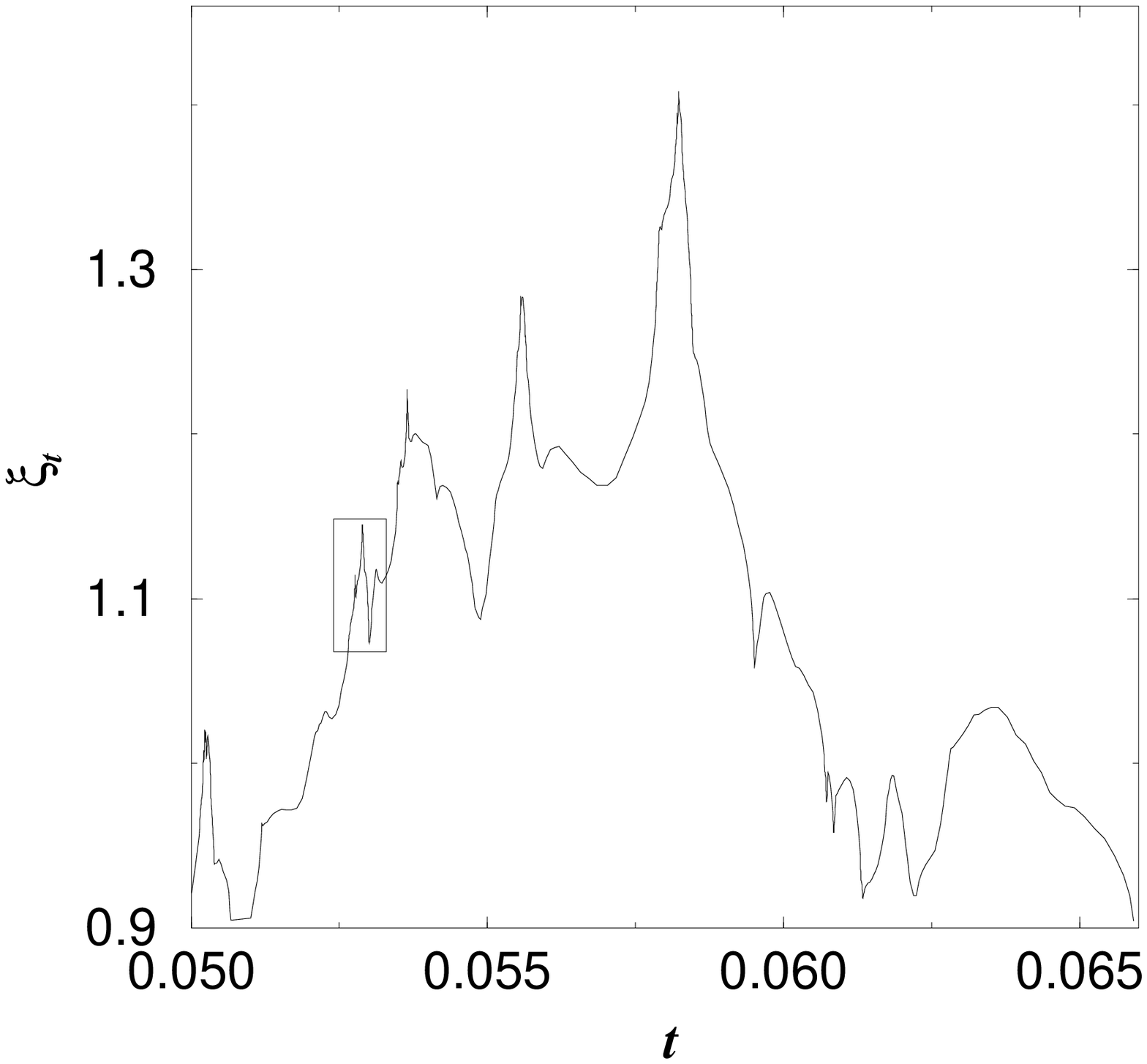}
      \vspace{1.ex}
      \centering (a)
    \end{minipage}\hfill
    \begin{minipage}[b]{.49\linewidth}
      \centering
      \includegraphics[angle=-90,width=.96\linewidth]{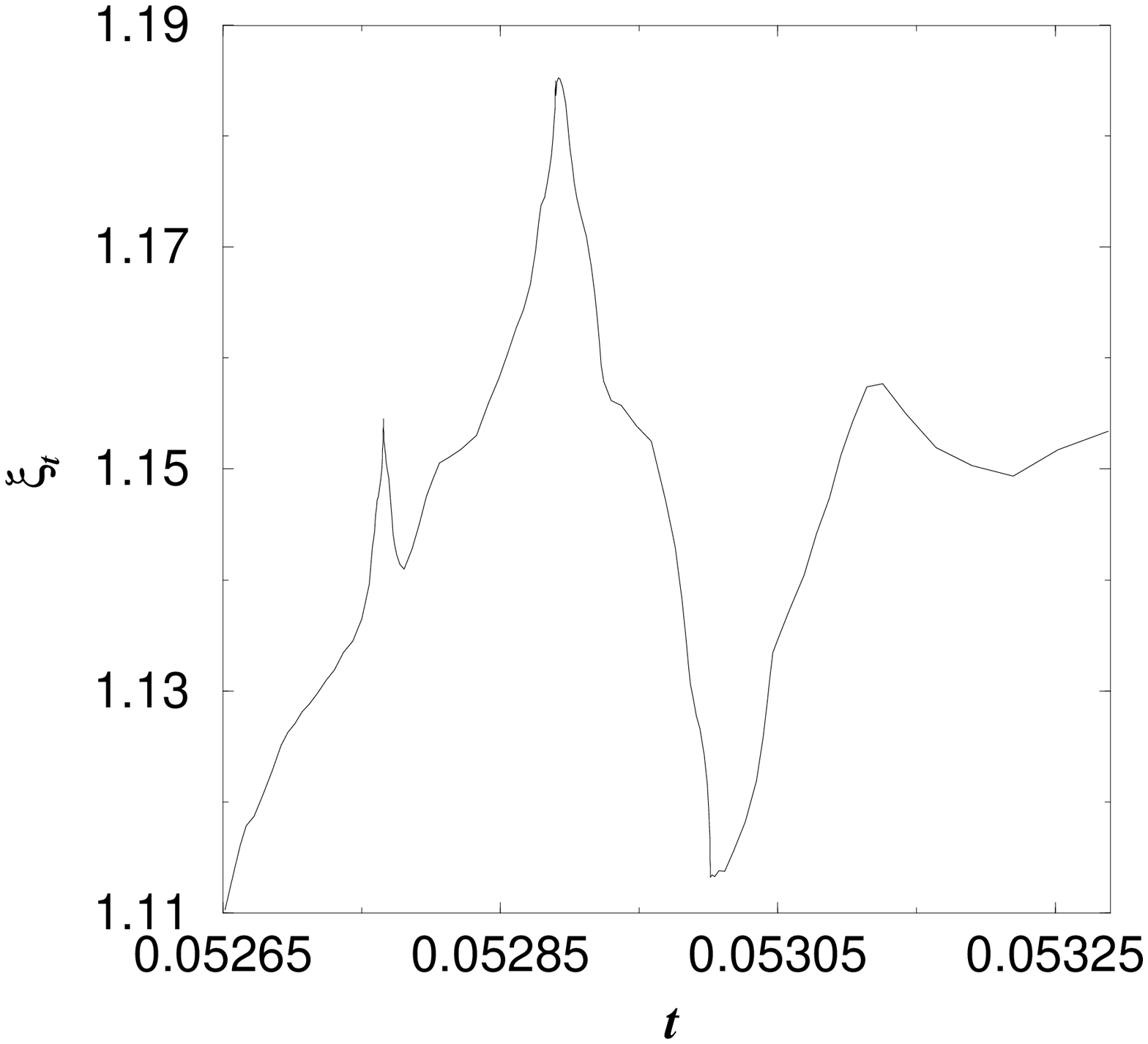}
      \vspace{1.ex}
      \centering (b)
    \end{minipage}
    \caption{(a)  Magnification of the small rectangular region in
      figure~\ref{forcing}. (b) Magnification of the rectangular region in
      (a).}  
    \label{forcing_details}
\end{figure}

In figure~\ref{forcing} we plot the forcing function which corresponds to the
nodal line of figure~\ref{exCh} and in figure~\ref{forcing_details} fine details
of this function are given.

We have numerically computed 2248 different realizations of the random wave
function (\ref{form_gen_fc}) with  $k=\sqrt{E}=100$. For each realization we
have found the nodal lines which were stopped when they hit the half-circle as
in figure~\ref{exCh}.  
 \begin{figure}[!b]
  \centering
  \includegraphics[angle=-90,width=.6\linewidth]{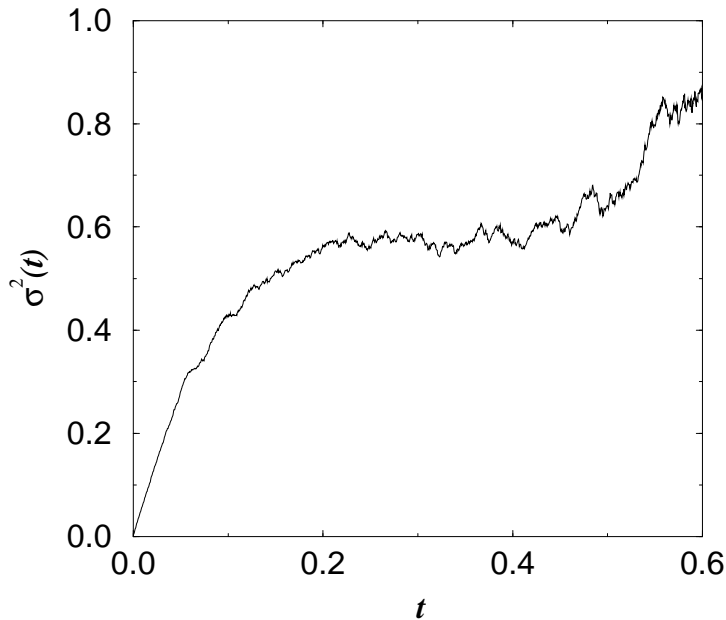}
  \caption{Variance of forcing functions for random waves inside the half-circle. }
  \label{variance_total}
\end{figure}

Then  corresponding  forcing
functions, $\xi_t(j)$, were calculated using the geodesic algorithm.  To check
that these functions are well described by the Brownian curves we calculated
mean value  $\bar{\xi}_t$  and  variance $\sigma^2(t)$ for each $t$   from the usual relations
\begin{equation}
\bar{\xi}_t=\frac{1}{N}\sum_{j=1}^N\xi_t(j)\; ,\;\;
\sigma^2(t)=\frac{1}{N}\sum_{j=1}^N(\xi_t(j)-\bar{\xi}_t)^2\;.
\label{statistics}
\end{equation}
In figure~\ref{variance_total} the dependence $\sigma^2(t)$ is plotted for all
$t$. It is clearly seen that it is not linear. 

This is not surprising as the linear behaviour is predicted only for infinite
curves.  Curves closed to boundaries which do not exist in SLE require a
special treatment which we will discuss in the next section. For short time
when curves do not feel the boundary the variance is close to be linear. 
 
\begin{figure}[!ht]
  \centering
  \includegraphics[angle=-90,width=.7\linewidth]{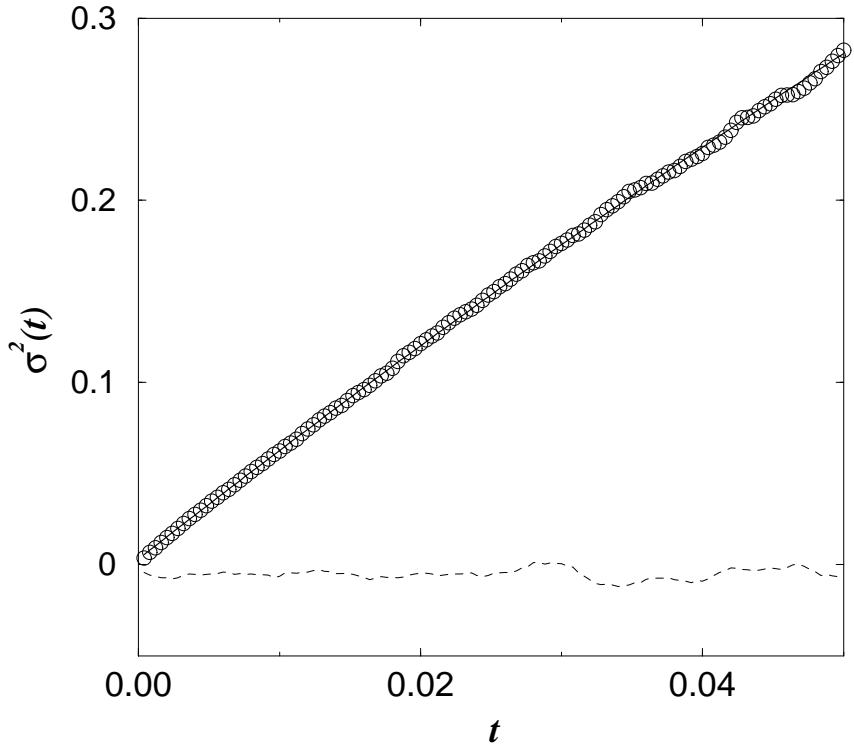}
  \caption{Variance (circles) and mean value (dashed line) in a short
    range. Solid line is the best quadratic fit (\ref{fit}).} 
  \label{var}
\end{figure}

In figure~\ref{var} the variance and the mean value are presented in a short
interval of $t$ where  $\sigma^2(t)$ is approximately linear. The solid line
in this figure is the best quadratic fit to the variance 
\begin{equation}
y=-0.003+6.05t-10.0t^2\;. 
\label{fit}
\end{equation}
The most important for us is the value of the linear term, $6.05$. It is
quite close to the pure percolation value  $\kappa=6$.  The mean value of the
forcing functions ($\approx 0$) also agrees with the SLE value.

\begin{figure}[!t]
 \centering
  \includegraphics[angle=-90,width=.7\linewidth]{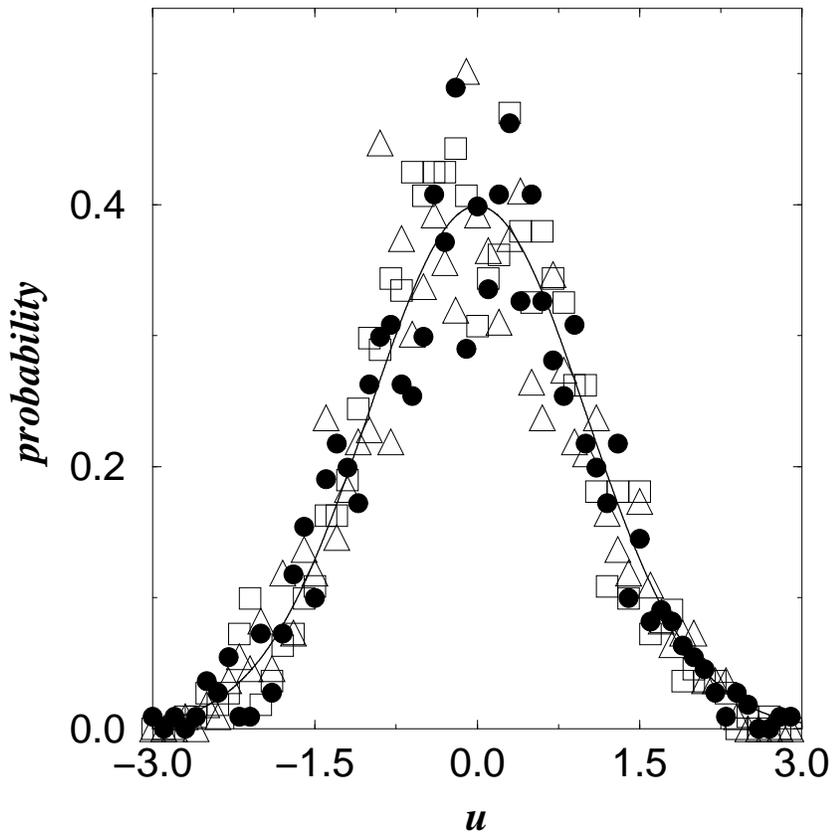}
  \caption{Probability law for random wave forcing functions. The abscissa
    axis is, as it follows for SLE$_6$,   
  $u=\xi_t/\sqrt{6t}$. The solid line is the standard Gaussian with unit variance: 
  $P(u)=\exp(-u^2/2)/\sqrt{2\pi}$. The points indicated by  $\fullcircle$,
  $\opentriangle$, and  $\opensquare$ correspond to, respectively,
  $t=0.02$, $0.04$, and $0.06$. }  
  \label{gauss}
\end{figure}

Another important prediction of the SLE$_6$ description of the percolation model
is that the distribution of values of $\xi_t$ with fixed $t$ has to be
Gaussian with variance $6t$. In figure~\ref{gauss} this distribution is
presented for different values of $t$. When the abscissa axis is rescaled by
$\sqrt{6t}$ all data are quite close to the Gaussian curve with unit
variance. 
Similar results were obtained when instead of the half-circle we consider
nodal lines inside a rectangle. 

The agreement of our results with SLE$_6$ predictions confirms that the nodal
lines of random wave functions are well described by the percolation model, as
was conjectured in \cite{charles}.   

\section{Dipolar SLE}\label{dipolar}

Though the results of the previous section are in a good agreement with
SLE$_6$,  let us notice two facts. First, the approximately linear increase
of the variance with time is observed only for a very small time
interval. Second, even within this interval a quadratic term in $t$ always
exists and is quite large (cf. (\ref{fit})). Both of these drawbacks are
related with inevitable  finite size effects and can be reduced by the using of 
another variant of SLE. 

The situation is analogous  to the Brownian motion in a finite interval. It
is well known (see e.g. \cite{feller}) that  the probability for an
one-dimensional Brownian particle starting from $x_0$ to arrive to the point $x$
at time $t$ in the whole space is given by the Green function of the
diffusion equation
\begin{equation}
G_0(x,x_0;t)=\frac{1}{\sqrt{2\pi Dt}}\exp \left ( -\frac{(x-x_0)^2}{2Dt}\right )
\end{equation}
where $D$ is a diffusion coefficient. Standard relations $<x>=x_0$ and
$<(x-x_0)^2>=Dt$ are simple consequences of this expression. 

For the Brownian motion inside a finite interval  such
probability depends on the boundary conditions. There are two main types of
boundary conditions: reflecting and absorbing. When a particle hits a
reflecting boundary it reflects back, when it touches an absorbing boundary
it stops. It is known (see e.g. \cite{feller}) within the formalism of the
diffusion equation that a reflecting boundary gives rise to the Neumann boundary
condition and an absorbing one corresponds to the 
Dirichlet condition. When both boundaries of an interval of length $h$ are
of the same type one gets that the Green function of this interval is
\begin{equation}
G_{\pm}(x,x_0;t)=\sum_{m=-\infty}^{\infty}\big(G_0(x+2mh,x_0;t)
 \pm G_0(-x+(2m+1)h,x_0;t) \big)\ \
\label{nd}
\end{equation}
Here $+$ (resp. $-$) denote Neumann (resp. Dirichlet) boundary conditions imposed at points $\pm h/2$.

\begin{figure}[!h]
 \centering
 \includegraphics[angle=-90, width=.6\linewidth]{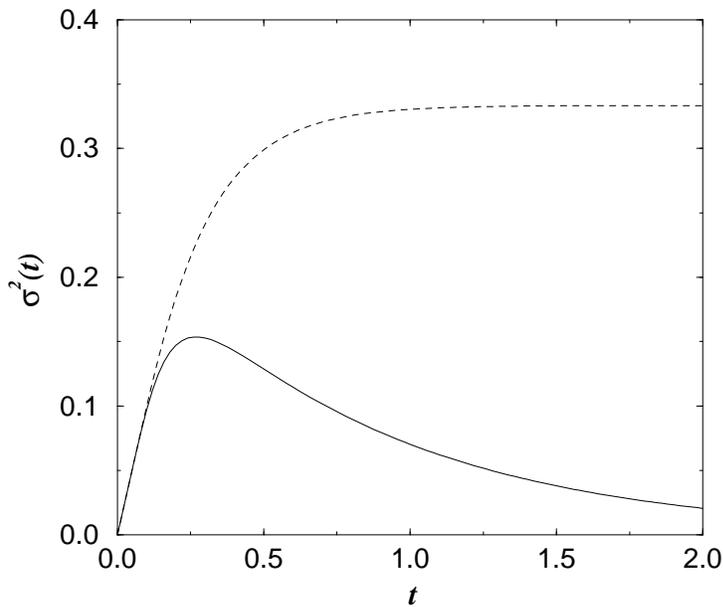}
  \caption{Variance for the one-dimensional Brownian motion with absorbing
    boundaries (solid line) and reflecting ones (dashed line). } 
  \label{bv}
\end{figure}

In figure~\ref{bv} the mean variance 
\begin{equation}
\sigma^2(t)=\int_{-h/2}^{h/2}(x-\bar{x})^2G_{\pm}(x,x_0;t){\rm d}x
\end{equation}
is plotted as a function of $t$ for the case $x_0=0$ and $D=1$. It is
clearly seen that the long time behaviour depends on the boundary conditions
and that the variance is linear in $t$ only for very small $t$ .    
These curves are similar to figure~\ref{variance_total} where the linear
behaviour was observed only for $t<0.05$ (cf. figure~\ref{var}).  We estimated
that in the best cases this interval corresponds to, roughly speaking, only
1/4 of the total curve length.  

To take into account larger parts of the curves and to check the
Brownian-type behaviour for a longer time interval it is  required to construct
formulae similar to (\ref{nd}) for SLE. 

The simplest case corresponds to the reflecting boundary conditions. In this
setting one considers a region and random curves (nodal lines) which start
at one fixed point, $z_0$, of the boundary and end at another fixed
boundary point, $z_1$. For the nodal lines this can be achieved e.g. by imposing
that  one boundary arc from $z_0$ to $z_1$ is positive and the second one is
negative. This type of processes is called chordal SLE (see
e.g. \cite{cardy}). By definition, it can be reduced to the standard SLE
from $0$ to infinity (cf. (\ref{loewner})) by a conformal transformation
which transforms the given region to the upper half plane in such a way that
the point $z_0$ is mapped to the origin and the point $z_1$ to infinity.   

Here, we would like to use a different variant of SLE which corresponds to a
region with two boundary arcs restricted by points $z_{+}$ and $z_{-}$. One
arc is assumed to be a reflecting boundary and the other is an absorbing
boundary. The random curves emerge from the point $z_0$ on the reflecting boundary
and are stopped when they hit the absorbing arc. 
By a conformal transformation our region can be transformed to the standard strip ${\mathbb{S}}$   
\begin{equation}
{\mathbb{S}}=\left\{ z \in \mathbb{C}, 0< \mathcal{I}\textrm{m } z <  \pi  \right\}\;
\label{strip}
\end{equation}
in such a way that points $z_{-}$, $z_+$, and $z_0$ are  mapped  to
$-\infty$,  $+\infty$, and $0$ respectively.      

In \cite{dipolar} it was shown that the SLE$_\kappa$ process which joins $0$
to  a point of the line $\mathcal{I}\textrm{m } z = \pi$ is described by the
following Loewner-type equation  
\begin{equation}
\frac{{\rm d} g_t(z)}{{\rm d} t} = \frac{2}{\tanh (g_t(z)-\xi_t)}
\label{dipolar_sle}
\end{equation}
where $g_0(z)=z$ and the forcing function, $\xi_t$,  is as above  the
standard Brownian motion (\ref{standard}) with variance $\kappa t$. 

To construct the dipolar SLE numerically from the data used in the previous
section we first transform the nodal lines inside the half-circle, as in
figure~\ref{exCh}, to the standard strip   
(\ref{strip}) by the following transformation
\begin{equation}
F(z)=\ln  [(L+z)^2/(L-z)^2 ]
\label{fz}
\end{equation}
where $L$ is the circle radius.
 
\begin{figure}[!t]
  \begin{minipage}[b]{.35\linewidth}
  \centering
  \includegraphics[angle=-90,width=.99\linewidth]{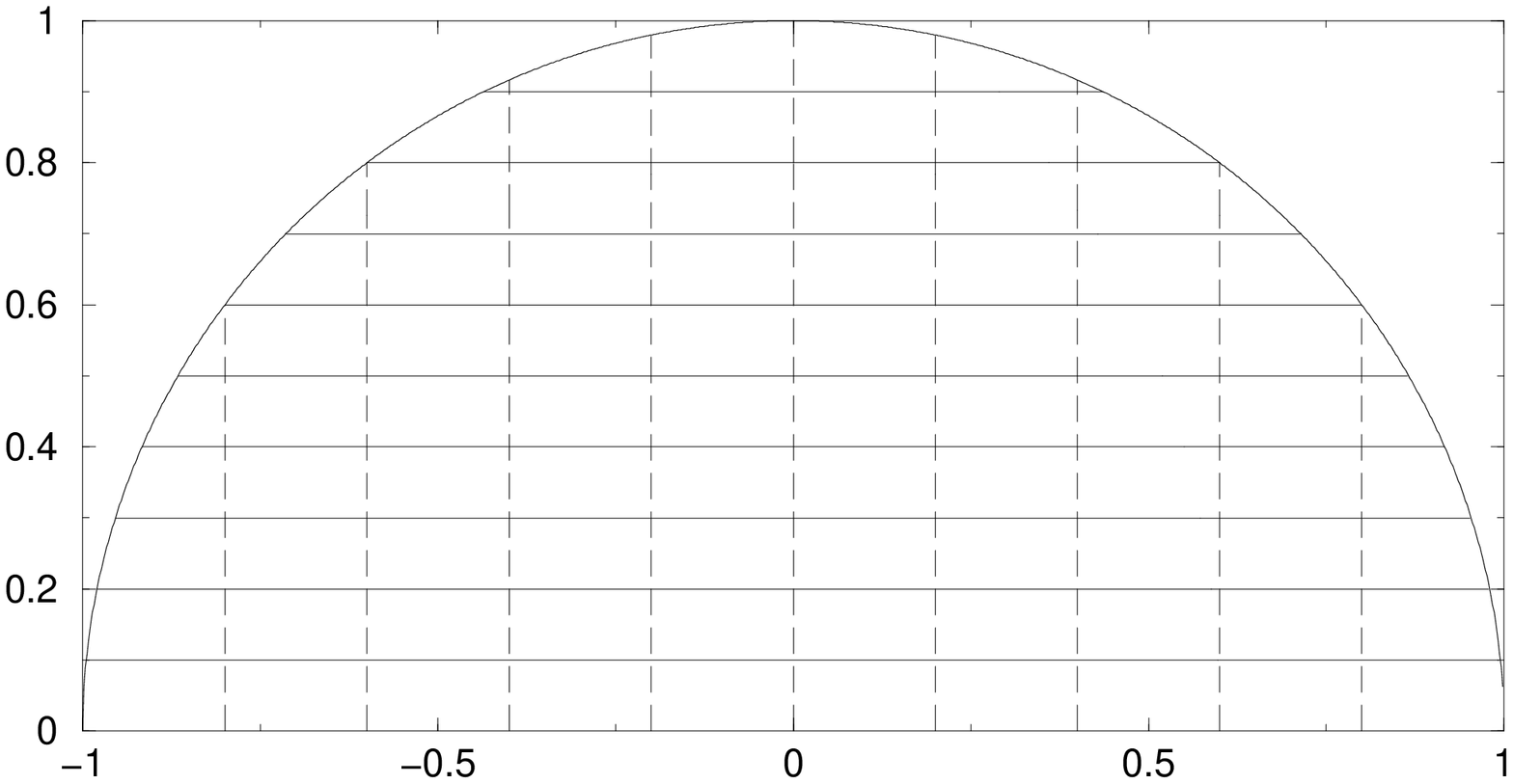}
  
  \vspace{1.ex}
  
  \centering (a)
  \end{minipage}
  \begin{minipage}[b]{.63\linewidth}
  \centering
  \includegraphics[angle=-90,width=.99\linewidth]{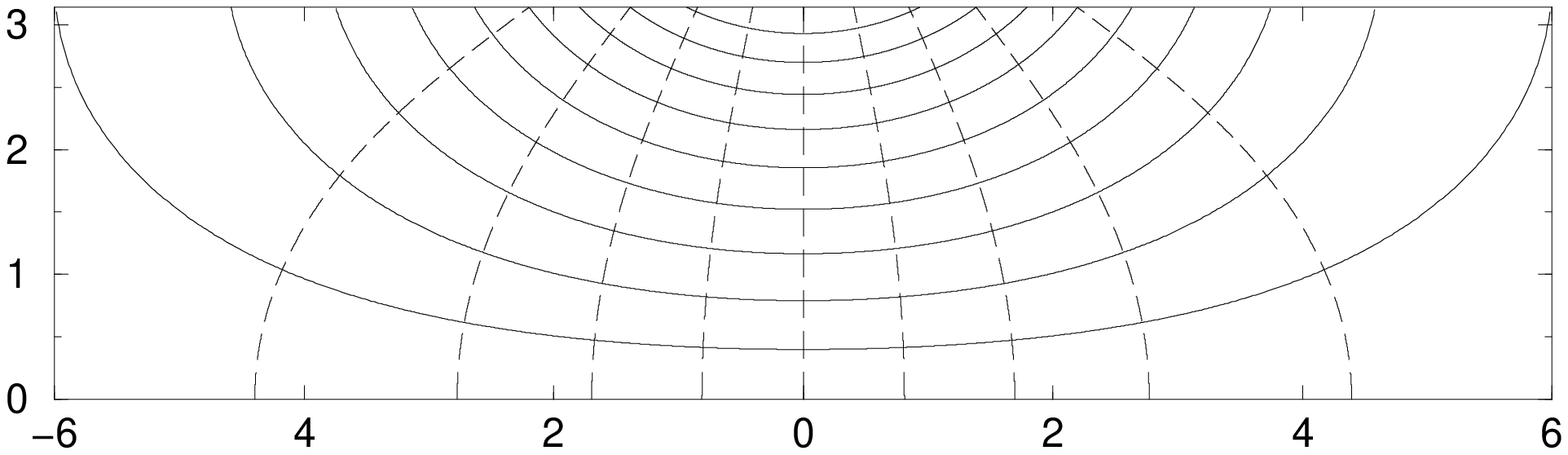}
  
  \vspace{1ex}
  
  \centering (b)
  \end{minipage}
  \caption{(a) Rectangular lattice inside the half-circle. (b) Its image under the map (\ref{fz}).}
  \label{circle_map} 
\end{figure}

To visualize this mapping  we present  the images of a rectangular lattice
inside the half-circle in figure~\ref{circle_map}. Note the strong
deformations of the regions close to the absorbing boundary.

\begin{figure}[!ht]
  \centering
  \includegraphics[angle=-90,width=.7\linewidth]{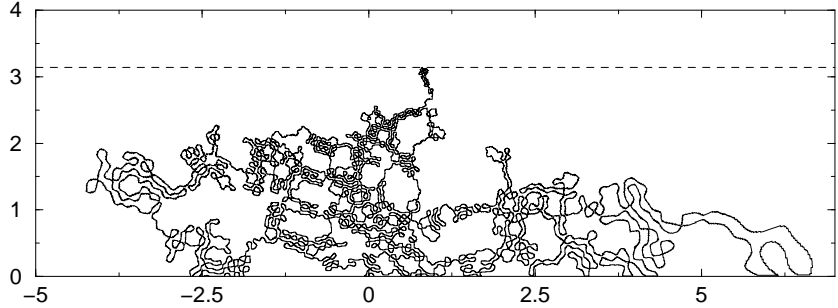}
  \caption{Image of the nodal line of figure~\ref{exCh} under the map
    (\ref{fz}). The dashed line is the absorbing boundary.}
  \label{trace_bande}
\end{figure}

The first step of calculations  consists in the transformation of every nodal lines inside the
half-circle to lines inside the chosen strip (\ref{strip}). Such an example
is illustrated in figure~\ref{trace_bande}.  

\begin{figure}[!b]
  \begin{center}
  \includegraphics[angle=-90,width=.7\linewidth]{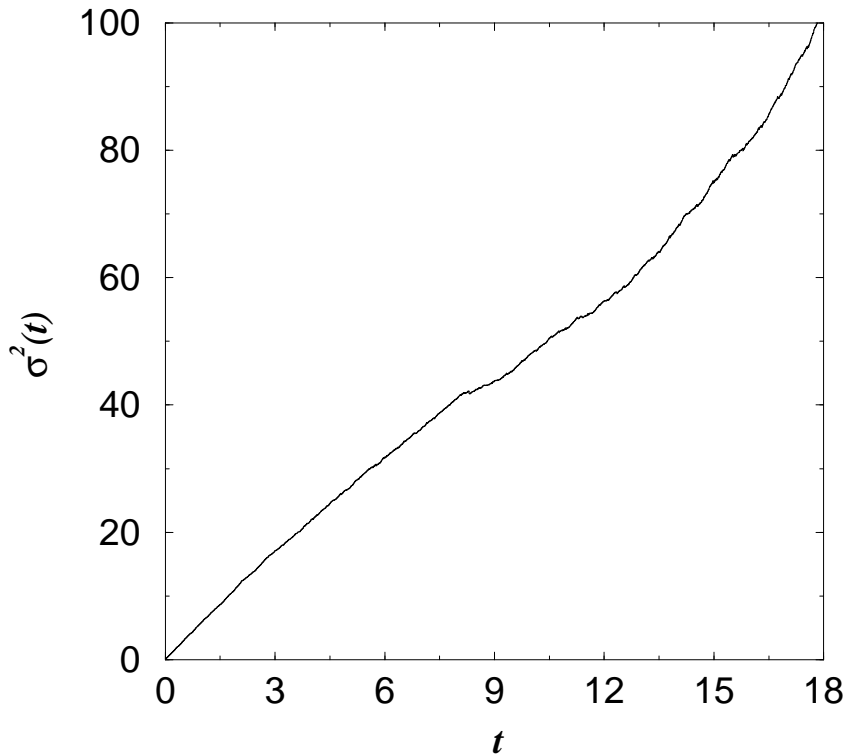}
  \caption{Variance of forcing functions for the nodal lines inside the strip.}
  \label{var_strip}
  \end{center} 
\end{figure}

\begin{figure}[!ht]
  \centering
  \includegraphics[angle=-90,width=.7\linewidth]{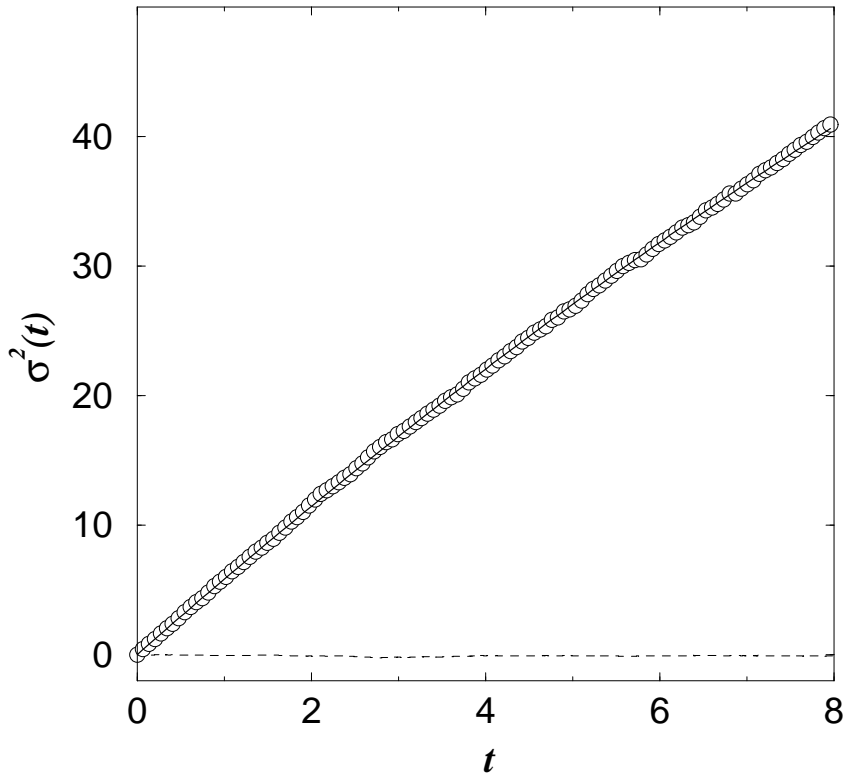}
  \caption{Variance (circles) and mean value (dashed line)  for 2252
    realizations inside the strip. Solid line: the best fit
    (\ref{fit_strip}).} 
  \label{var_strip_small}
\end{figure}

Then it is necessary to find the dipolar conformal transformation which map the strip with
a line to the strip itself (cf. (\ref{dipolar_sle})).  In the appendix  a simple
algorithm for such a mapping is briefly discussed. Using it we compute
numerically the forcing function, $\xi_t$ for each nodal line and calculate
its statistical properties. In figure~\ref{var_strip} the variance
(\ref{statistics}) is plotted for all curves and in
figure~\ref{var_strip_small}  the region of linear increase of $\sigma^2(t)$
is magnified. The solid line in this figure indicates the best quadratic fit 
\begin{equation}
y=5.92t-0.103t^2\;.
\label{fit_strip}
\end{equation}  
The coefficient  $5.92$ is to compare with the percolation theory value of $\kappa=6$. 
Other statistical characteristics are also in a good agreement with the SLE$_6$ predictions.

Contrary to (\ref{fit}) the quadratic term in (\ref{fit_strip}) is small.
It is mainly related with discretization errors and it decreases when more
points along a curve are taken into account. The non-linear behaviour of the
variance in figure~\ref{var_strip} for large $t$ is connected with large
errors due to the stretching of lines close to the boundary as it is evident
from  figures~\ref{circle_map} and \ref{trace_bande}. We checked that the
interval $t< 8$  in figure~\ref{var_strip_small} where the variance is linear in
$t$ corresponds approximately  to $92\%$  of the total curve length.  It means
that  the dipolar SLE$_6$ is a good description of the nodal lines practically
for the entire curves. Exceptions are the small line parts close to the absorbing 
boundary which require better approximations.  

We have also investigated nodal lines inside a rectangle $[-L/2;L/2]\times
[0;l]$ which can be conformally mapped into the strip (\ref{strip}) by the
following transformation 
\begin{equation}
F(z)=-\ln [(\wp(z+L/2)-\wp(L))/(\wp(L/2)-\wp(L))]  
\end{equation}
where $\wp(z)$ is the Weierstrass elliptic function with periods $2L$ and
$2{\rm i} l$ (see e.g. \cite{bateman}).  

The results in this case are similar to the ones above and all agree well
with the percolation model. 

\section{Conclusions}\label{conclusion}

In summary, we investigated the nodal lines of two-dimensional random wave
functions by numerically computing one-dimensional forcing functions of
Loewner's evolution. We demonstrate that the later are well described by a
Brownian motion with zero mean and variance growing linearly in time. The
coefficient of proportionality is close to the value $6$ predicted by the
percolation model. Our results give an additional support to the conjecture
that the nodal domains of random wave functions in the scaling limit are described
by the critical percolation.  We show that using dipolar SLE
reduces significantly finite size effects.   

\section*{ Acknowledgements}

The authors are very thankful to J. Keating and I. Williams for useful
discussions and, in particular, for pointing out an error in the
calculations. One of the authors, E.B., is grateful to G. Falkovich for
a discussion of the paper \cite{turb} prior the publication and to D. Bernard
for  clarifying discussions about SLE.   

\section*{Appendix}
Consider a simple curve which starts on the real axis and grows inside a
region of  the upper half plane. The goal of the numerical algorithms below is to
find a forcing function in (\ref{loewner}) corresponding to a conformal map
which transforms the region cut along the curve into the region itself.  

Let $(x_0,0)$ be coordinates of the first point of the curve which we assume
belongs to the abscissa axis and let $(x_1,y_1)$ be coordinates of the next
point closest to the first.  
The simplest numerical method consists of assuming that the forcing function
is a piecewise constant function.  The map with a constant forcing function
$\delta \xi$ for $0<t\leq \delta t$  is given by (\ref{slit}) as
\begin{equation}
g_t(z)=\delta \xi+\sqrt{(z-\delta \xi)^2+4\delta t}\;.
\label{a_slit}
\end{equation}
The parameters $\delta \xi$ and $\delta t$ are obtained from  the condition that
the  top of the slit coincides  with the point $\delta z=x_1-x_0+{\rm i}
y_1$, which yields
\begin{equation}
\delta \xi={\cal R}e \, \delta z\;,\;\;\;\delta t=\frac{1}{4}({\cal I}m\, \delta z)^2\;.
\label{pw}
\end{equation}
Then we transform all points of the curve except the first one by (\ref{a_slit})
with these values of parameters and renumber the resulting points. One repeats
this process till the whole curve will be transformed. In \cite{bauer} it is
proved that such an algorithm converges for sufficiently small $\delta t_j$ and
$\delta x_j$.  

In section~\ref{chordal} we used the more refined geodesic algorithm
\cite{algorithm} in which one approximates a small part of the curve between
two points, $0$ and $\delta z$,  by the geodesic arc (a circle perpendicular
to the real axis) passing through these points.   Normalizing the map given
in \cite{algorithm} such as to obey the convention (\ref{normhy}) one gets
the expression  
\begin{eqnarray}
g_t(z)&=&\frac{b^3}{\sqrt{b^2+c^2}\sqrt{(bz/(b-z))^2+c^2}-b^2-c^2}+%\nonumber\\
\frac{2b^3+3bc^2}{2(b^2+c^2)}
\label{geod}
\end{eqnarray}
where
\begin{equation}
b=\frac{|\delta z|^2}{{\cal R}e \, \delta z}\;,\;\;\;c=\frac{|\delta z|^2}{{\cal I}m\, \delta z}\;.
\end{equation} 
Direct calculations give that the time corresponding to $\delta z$ and the
value of the forcing function in this point  are 
\begin{equation}
\delta t=\frac{1}{4}({\cal I}m\, \delta z)^2+\frac{1}{8}({\cal R}e \, \delta z)^2
\label{geod_1}
\end{equation}
and
\begin{equation}
\delta \xi_t=\frac{3}{2}{\cal R}e \, \delta z\;.
\label{geod_2}
\end{equation}
When a point moves along a geodesic circle of diameter $b$ the forcing function changes as   follows
\begin{equation}
\xi_t=\frac{12t}{b+\sqrt{b^2-8t}}
\end{equation}
and is practically linear for short time.  
Eqs.~(\ref{geod})--(\ref{geod_2}) define the geodesic algorithm.

To do some computations following dipolar SLE in Section~\ref{dipolar} we  use a piecewise
constant approximation for  the forcing function in (\ref{dipolar_sle}). In
this case one has the exact solution \cite{dipolar} 
\begin{equation}
\cosh \left ( \frac{1}{2}(g_t(z)-\delta \xi)\right )={\rm e}^{\delta
  t/2}\cosh\left (\frac{1}{2}(z-\delta \xi)\right )\;. 
\end{equation}
Finding $\delta t$ and $\delta \xi$ from the condition that the curve  tip
coincides with $\delta z$ one obtains   
\begin{equation}
\delta \xi= {\cal R}e \, \delta z\;,\;\;
\exp \left (-\frac{1}{2}\delta t\right )=\cos \left (\frac{1}{2}{\cal I}m\,
  \delta z\right )\;. 
\end{equation}
These expressions are the dipolar analog of Eqs.~(\ref{pw}) and they permit to
construct the simplest algorithm of numerical calculations for the dipolar
case.

\end{document}